\documentclass{pasa}%

\title[Scientific Goals of KISS]{Scientific Goals of the Kunlun Infrared Sky Survey (KISS)}
\author[Michael Burton et al.]{Michael G. Burton$^{1,2}$, Jessica Zheng$^{3}$, Jeremy Mould$^{4,5}$, Jeff Cooke$^{4,5}$, Michael Ireland$^{6}$, Syed Ashraf Uddin$^{7}$, Hui Zhang$^{8}$, Xiangyan Yuan$^{9}$, Jon Lawrence$^{3}$, Michael C.B. Ashley$^{1}$, Xuefeng Wu$^{7}$, Chris Curtin$^{4}$ \and Lifan Wang$^{7,10}$ \\
\affil{$^1$School of Physics, University of New South Wales, Sydney, NSW 2052, Australia}
\affil{$^2$Armagh Observatory and Planetarium, College Hill, Armagh, BT61 9DG, Northern Ireland, UK}
\affil{$^3$Australian Astronomical Observatory, 105 Delhi Road, North Ryde, NSW 2113, Australia}
\affil{$^4$Centre for Astrophysics and Supercomputing, Swinburne University of Technology, PO Box 218, Mail Number H29, Hawthorn, VIC 3122,  Australia}
\affil{$^5$ARC Centre of Excellence for All-sky Astrophysics (CAASTRO)}
\affil{$^6$Australian National University, Canberra, Australia}
\affil{$^7$Purple Mountain Observatory, Chinese Academy of Sciences, 2 West Beijing Road, Nanjing, 210008, China}
\affil{$^8$School of Astronomy and Space Science, Nanjing University, Nanjing, China}
\affil{$^9$Nanjing Institute of Astronomical Optics \& Technology, Chinese Academy of Sciences, 188 Bancang Street, Nanjing 210042, China}
\affil{$^{10}$Texas A\&M University, Texas, USA}
~\\
Accepted for Publication in PASA, 04/08/16.
~\\
}
\jid{PASA}
\doi{10.1017/pas.\the\year.xxx}
\jyear{\the\year}

\usepackage{aas_macros}
\usepackage{hyperref} 
\hypersetup{colorlinks,citecolor=blue,linkcolor=blue,urlcolor=blue}

\begin{document}%
\begin{abstract}
The high Antarctic plateau provides exceptional conditions for conducting infrared observations of the cosmos on account of the cold, dry and stable atmosphere above the ice surface.  This paper describes the scientific goals behind the first program to examine the time-varying universe in the infrared from Antarctica -- the Kunlun Infrared Sky Survey (KISS).  This will employ a small (50\,cm aperture) telescope to monitor the southern skies in the $\rm 2.4\mu m\, K_{dark}$ window from China's Kunlun station at Dome A, on the summit of the Antarctic plateau, through the uninterrupted 4-month period of winter darkness.  An earlier paper discussed optimisation of the $\rm K_{dark}$ filter for the best sensitivity \cite{2016PASA...33....8L}.  This paper examines the scientific program for KISS\@.  We calculate the sensitivity of the camera for the extrema of observing conditions that will be encountered.  We  present the parameters for sample surveys that could then be carried out for a range of cadences and sensitivities.  We then discuss several science programs that could be conducted with these capabilities, involving star formation, brown dwarfs and hot Jupiters, exoplanets around M dwarfs, the terminal phases of stellar evolution, discovering fast transients as part of multi-wavelength campaigns, embedded supernova searches, reverberation mapping of active galactic nuclei, gamma ray bursts and the detection of the cosmic infrared background.
\end{abstract}
\maketitle%
\section{INTRODUCTION}
\label{sec:intro}

The Kunlun Infrared Sky Survey (KISS) will be conducted from China's Kunlun station at Dome A, at the summit of the Antarctic plateau. The survey will be undertaken in the K--dark passband at 2.4$\mu$m, a unique low background window in the infrared. It will be conducted with the third of the AST--3 telescopes (Antarctic Survey Telescope) deployed at Dome A \cite{2008SPIE.7012E..2DC}.  The first two telescopes were designed for optical surveys.  The third \cite{spie2016} will conduct the first deep, wide-field survey in this band, as well as be pathfinder for a future large Antarctic optical/IR telescope at the site (KDUST).  This paper discusses the science objectives for the infrared surveys that will be conducted with KISS\@.

Sky surveys are an essential prerequisite to astrophysics. To understand the physics of the Universe, you need to know what is there. That is why, for example, the UK Schmidt telescope was built at Australia's national Siding Spring Observatory, ahead of the Anglo Australian Telescope in the 1970s \cite{1985QJRAS..26..456L}. KISS will be complementary to the ANU's SkyMapper telescope \cite{2007PASA...24....1K} in that it is infrared and complementary to NASA's 2MASS survey \cite{2006AJ….131.1163S} in that it is time sensitive. The facility uniquely provides complete southern hemisphere sky coverage in the infrared with surveys with cadences that may range from hourly to monthly. The AST3--NIR project is a precursor for a KDUST IR camera \cite{2013IAUS..288..239M, 2014SPIE.9145E..0EZ}. Just as KISS complements SkyMapper, KDUST will complement the optical Large Synoptic Survey Telescope (LSST), the top ranked ground based facility in the U.S. Astro2010 decadal plan\footnote{See http://sites.nationalacademies.org/bpa/bpa\_049810.}. KISS also allows Chinese astronomers to acquire first hand near--IR data of the southern sky, as well as produce a range of astronomical sources which will stimulate further collaboration between Chinese and Australian scientists. 

KISS exploits three particular advantages that the Antarctic plateau has for astronomical observations over other ground-based locations:

\begin{itemize}
\item The low sky background at 2.4$\mu$m, two orders of magnitude lower than at the best temperate sites \cite{1996PASP..108..721A,1996PASP..108..718N}.
\item The high photometric precision achievable due to low scintillation and the exceptional seeing above narrow surface boundary layer \cite{2004Natur.431..278L,2006PASP..118..924K,2010PASP..122.1122B,2010AJ....140..602Z,2016AJ....151..166O}.
\item The high time cadence that can be achieved through the 4-month darkness of a winter night \cite{2015ApJS..218...20W,2015AJ....149...25H}.
\end{itemize}

2.4$\mu$m is the longest wavelength that truly deep imaging can be undertaken from the Earth.  It marks the outer limit for optical astronomy, beyond which thermal background fluxes dominate observations. 

In this paper we describe the KISS instrument in \S\ref{sec:instrument}, calculate its sensitivity in \S\ref{sec:sensitivity} and describe its operation and automated data reduction pipeline in \S\ref{sec:operations}.  We then discuss the survey capabilities in \S\ref{sec:survey}, before outlining a range of science programs in \S\ref{sec:science} that will be undertaken.  These include exoplanet and brown dwarf searches, studies of the accretion process during star formation, of the terminal phases of stellar evolution and investigations of fast transients across multi-wavelengths.  We also discuss searches for buried SN, reverberation mapping of AGN nuclei and the measurement of the cosmic infrared background.

\section{THE INSTRUMENT}
\label{sec:instrument}

The Antarctic Survey Telescope--AST3 was originally designed to have three optical telescopes matched respectively with G, R and I band filters, and used principally in searches for supernovae and exosolar planets undertaken from Dome A\@. Each of the three AST3 telescopes is a modified Schmidt type optical telescope with an entrance pupil diameter of 500\,mm, an $f$-ratio of 3.7 and a field of view of $4.1^{\circ}$. They consist of a transparent aspherical plate as the entrance pupil, an aspherical primary mirror with a diameter of 680\,mm, and a spherical refractive corrector with a filter before the focal plane \cite{2012MNRAS.424...23Y}. The first two telescopes were installed at Dome A in January 2012 and 2015, respectively \cite{2012SPIE.8444E..1OL,2014SPIE.9145E..0FY}.

Due to the great advantages of Dome A for near--IR observations, the third AST3 telescope (AST3--NIR $\equiv$ KISS) has been re-optimized for diffraction limited imaging in the K band \cite{spie2016}, based on the same entrance pupil and primary mirror as that of the first two AST3 telescopes, since they have already been manufactured at the NIAOT\@. This results in a primary mirror which is oversized for the entrance pupil for the camera.  There will be a single filter centred at $\lambda = 2.375\mu$m with bandwidth $\Delta \lambda = 0.25\mu$m; i.e.\ $\rm K_{dark}$ \cite{2016PASA...33....8L}.  The KISS camera is located outside the main optical tube by adding in a fold mirror. The telescope is being built at the NIAOT and the infrared camera at the AAO\@. 

The telescope and camera specifications are listed in Table~\ref{tab:specs}.  The optical design is described in more detail in a separate paper \cite{spie2016}.

\begin{table*}
\caption{Specifications for the AST3 Telescope and the KISS IR Camera.}
\begin{center}
\begin{tabular}{@{}lcl}
\hline\hline
Parameter & Value & Comment \\
\hline%
Diameter of Primary, $D_{primary}$ & 68\,cm  & Set by the AST3 optical telescopes design \\
Entrance Window Pupil Diameter, $D$ & 50\,cm & Set by the KISS camera \\
Focal Ratio, $f/\#$ & 5.5 & Averaged over camera FOV \\
Array Size & $2048 \times 2048$ & Teledyne Array \\
Well Depth & 80,000 e  &  \\
Pixel Size & $18\mu$m \\ 
Pixel Scale, $\theta_{pixel}$ & $1.35''$  & From focal length\\
Optics Tolerance & $0.68''$ \\
Diffraction Size & $1.0''$ &  FWHM for pupil size \\
Seeing & $0.5'', 1.0'' \,\&\, 2.5''$ & Excellent, average \& poor conditions \\
Image Quality & $1.3'', 1.6''\, \&\, 2.8''$ & Diffraction + Seeing + Optics Tolerance \\
& $\equiv 1.1, 1.2\, \&\, 2.1$ &  times the diffraction Size \\
Camera Field of View & $46' \times 46'$ \\
Central Wavelength, $\lambda$ & $2.375\mu$m & $\rm K_{dark}$ \\
Bandwidth, $\Delta \lambda$ & $0.25\mu$m \\
Filter Response, $T_{filter}$ & 0.95 &  \\
Optical Efficiency of Telescope, $\tau_{tel}$ & 0.66 &\\
Optical Efficiency of Camera, $\tau_{cam}$ & 0.8 & \\
Quantum Efficiency of Detector, $\eta$ & 0.8 &  \\
Dark Current, $I_D$ & $< 1$ e/s/pix  &  \\
Read-out Noise, $R$ & $<3$\,e/pix,  $<18$\,e/pix  & Fowler, CDS read-out \\
Minimum Integration Time, $t_{min}$ &  2\,s & \\
\hline\hline
\end{tabular}
\end{center}
\label{tab:specs}
\end{table*}

\section{SENSITIVITY}
\label{sec:sensitivity}

Since sky backgrounds in the $\rm 2.4\mu m\, K_{dark}$ band at Dome A are much lower than from any temperate sites on the Earth, this gives the AST3--3 telescope a significant sensitivity advantage over similarly sized-telescopes located elsewhere. This advantage is particularly striking when also considering the telescope's highly stable, near diffraction-limited performance over a wide field of view, although the IR camera is still operated in a seeing limited mode. 

For point source observations, the sensitivity of the instrument can be presented as the integration time required to achieve a particular signal to noise ratio for a given observed object magnitude, spread over several pixels.  For extended source observations, we calculate the sensitivity per pixel and then convert to per square arcsecond.

To calculate the signal to noise ratio for the instrument, a simplified model for the background noise on the detector has been adopted.  We assume that the final image frame from detector is processed with accurate calibration using appropriate flat and dark frames. The following noise sources (all assumed Poisson distributed) are considered to contribute to the signal-to-noise calculation:

\begin{itemize}
	\item photon noise on the signal ($S$, measured in $\rm e/s$) from the object,
	\item photon noise on the background flux ($B$, measured in $\rm e/s/pix$) from the sky,
	\item photon noise on the thermal self emission ($TSE$, measured in in $\rm e/s/pix$) from the telescope and IR camera,
	\item read-out noise $R$ of the detector (measured in $\rm e/pix$), and
	\item shot noise on the dark current ($I_{D}$, measured in $\rm e/s/pix$).
\end{itemize}

The signal to noise ratio can found from equation \ref{eqn:SN}, which is adapted from McCaughrean \& McLean \cite{1987PhDT........14M,2008eiad.book.....M}: 

\begin{equation} \label{eqn:SN}
\frac{S}{N} = \frac{S \, \sqrt{n_{0}  \, t}} {\sqrt{S+n_{pixel} \, (B+TSE+I_{D}+ R^{2}/t)}}.
\end{equation}

Here $t$ is the integration time per exposure and $n_{0}$ the number of exposures (and is usually just one). 

$n_{pixel}$ is the number of pixels sampling the stellar image and can be estimated following Eqns.~2--4 of Li et al.\ \cite{2016PASA...33....8L} involving the pixel size, $\theta_{pixel}$ and the full width at half maximum intensity of the point spread function (PSF), $\theta_{FWHM}$. This latter quantity is dictated by the assumed seeing (which will typically be $\sim 1.0''$ but may be as good as $\sim 0.5''$ \cite{2010AARv..18..417B,2016PASA...33....8L}), the diffraction limit ($1.0''$) and the tolerance of the optics ($0.7''$). Note that the last two quantities are defined by the optical design of the KISS camera.  $\theta_{pixel}$ is the pixel scale of the system ($1.35''$).  
We also round $n_{pixel}$ to the nearest integer.  Hence, in the calculations presented here we have taken $n_{pixel}$ as 9 (i.e.\ $3 \times 3$ pixels) when the seeing is $0.5''$ and 25 when the seeing is $2.5''$.



For extended emission, we set $n_{pixel} = 2$ (so as to include 1 sky pixel for each source pixel), and then add $\rm 2.5 \log(\theta_{pixel}^2)$ to derive sensitivities in magnitudes per square arcsecond. 

When the telescope is operated in sky-background limit conditions, as it would ideally be, then the S/N is given by:

\begin{align}
\label{eqn:Backgroundlimited}
\frac{S}{N} = S \, \sqrt{\frac{n_{0}  \, t} {n_{pixel} \, B}}.
\end{align}

Conversely, for short exposures it may be readout noise limited, when the S/N would be:

\begin{align}
\label{eqn:Detectorlimited}
\frac{S}{N} = \frac{S \, t}{R} \sqrt{\frac{n_{0}}{n_{pixel}}} .
\end{align}

For the parameters of the KISS camera this provides the dominant noise source for integrations of less than 4\,s in duration.

The number of photons collected from an object of apparent magnitude of $m$, transmitted through the atmosphere with transmission of $T_{Sky}(\lambda)$, by a telescope of area $A_{tel}$, in a wavelength range $\Delta \lambda$ centred at $\lambda$ with filter response of $T_{filter}(\lambda)$, passing through an optical system of efficiency $\tau(\lambda)$, and measured by a detector with quantum efficiency of $\eta(\lambda)$, are then determined.

The details of this calculation can be found in our recently published paper \cite{2016PASA...33....8L}.  It is obvious that the sensitivity of the IR camera is related to the spectral properties in all parts in the optical path.  When a system is designed,  normally the spectrum of the sky transmission, the sky emission and the quantum efficiency of the CCD detector is defined. The filter performance  also contributes to the final sensitivity. This optimization has been discussed in our previous work in considering the telescope thermal self emission. An exposure time calculator has been developed and can be accessed through the following link: \url{http://newt.phys.unsw.edu.au/~mcba/ETC_WebModel/TELESCOPE.html}.

\begin{figure*}
	\begin{center}
		\includegraphics[width=2\columnwidth]{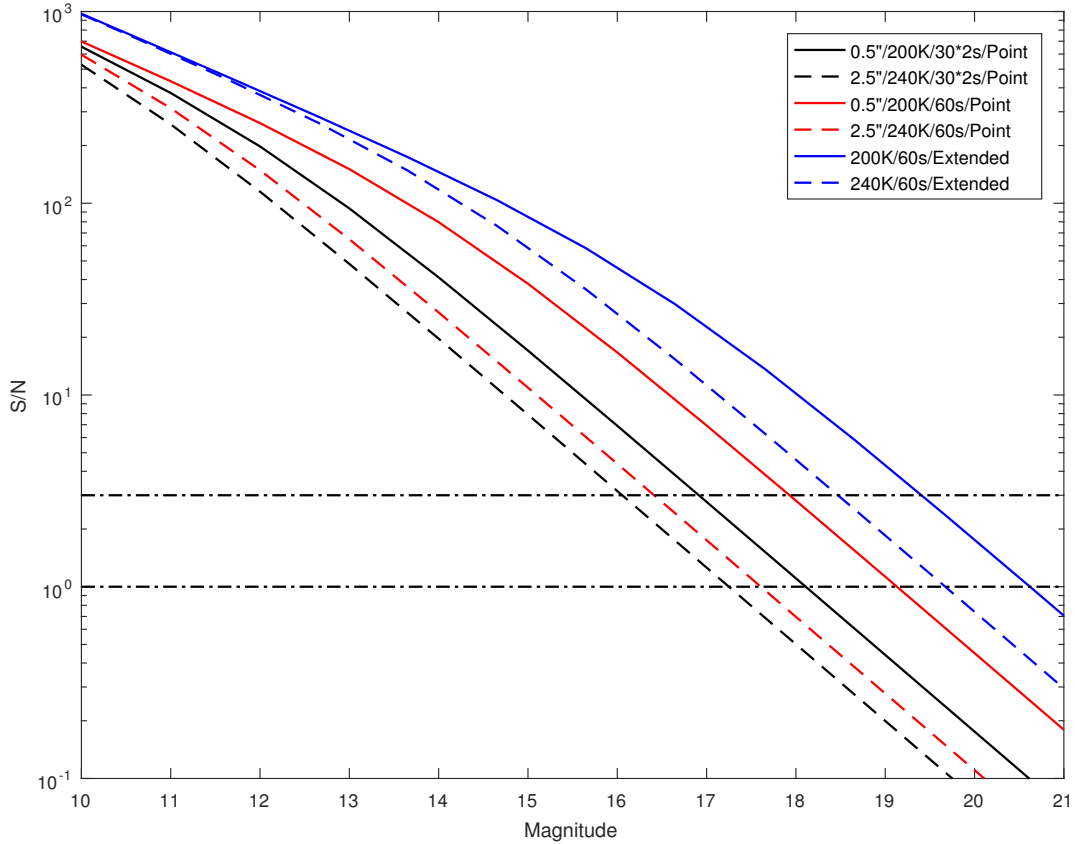}
		\caption{Signal to Noise {\it vs.} $\rm K_{dark}$ Vega magnitude achieved for a point source in two different seeing and ambient temperature conditions (solid lines for $\rm 0.5'', \,200\,K$ and dashed lines for $\rm 2.5'', \,240\,K$), as listed in the legend.  Red lines are for a single 60s exposure (i.e.\ in background limited operation), black lines are for $\rm 30 \times 2\,s$ exposures (i.e.\ when minimising saturation from bright stars). The S/N for extended source emission is shown with the blue lines (solid for 200\,K and dashed for 240\,K), though note that the flux units are now in magnitudes per square arcsecond.  The two horizontal dotted lines give the threshold for a S/N of 1 and 3, respectively.}
		\label{fig:SNvsMag}
	\end{center}
\end{figure*}

Figure \ref{fig:SNvsMag} shows the KISS camera signal to noise ratio (S/N)  {\it vs\@.} $\rm K_{dark}$ magnitude in a 60\,s exposure time for two different ambient temperature and seeing conditions, assuming a sky background of 16.9 mag in $\rm K_{dark}$ (i.e.\ about $\rm 100\mu mJy/arcsec^2$ \cite{1996PASP..108..721A}). The two conditions shown here represent the best and worst expected for the site.  While longer exposures than 1 minute might be taken to minimise the contribution due to read noise, sky stability may then limit the quality of the background subtraction.

It can be seen that when the ambient temperature is 200\,K with $0.5''$ seeing (corresponding to excellent winter-time conditions), the S/N is considerably better than when operated on poorer winter days (e.g.\ a warmer 240\,K and poor seeing of $2.5''$). The difference between these two extreme conditions is $\sim 1.5$ magnitudes in sensitivity for the same S/N\@. 

The Figure also shows the S/N when the integration uses $\rm 30 \times 2s$ frames, instead of a single $\rm 60s$ frame.  This would be the case if saturation of bright stars is to be minimised in a program.  This then increases the dynamic range between the brightest and faintest stars that could be measured simultaneously.  It does, of course, lead to a decrease in the S/N as measurements will no longer be background limited.  As can be seen in Fig.~\ref{fig:SNvsMag}, the sensitivity loss is $\sim 0.4 - 1.0$ mags for fainter stars (dependent on the conditions), but negligible for bright stars.

The S/N for an extended source in a single 60\,s exposure is also shown in Fig.~\ref{fig:SNvsMag} vs.\ the brightness (now in magnitudes per square arcsecond), for the same two ambient temperatures (200\,K \& 240\,K).  The sensitivity difference in this case is about 1 mag.\ for the same S/N\@.

In general, nine jittered observations will be needed to remove stars to derive sky and flat field frames for each source position, so that the sensitivity achieved would be 3 times better than shown in Fig.~\ref{fig:SNvsMag} for a targeted observation of a field, albeit for a 9 times longer observation time.

\begin{figure*}
	\begin{center}
		\includegraphics[width=2\columnwidth]{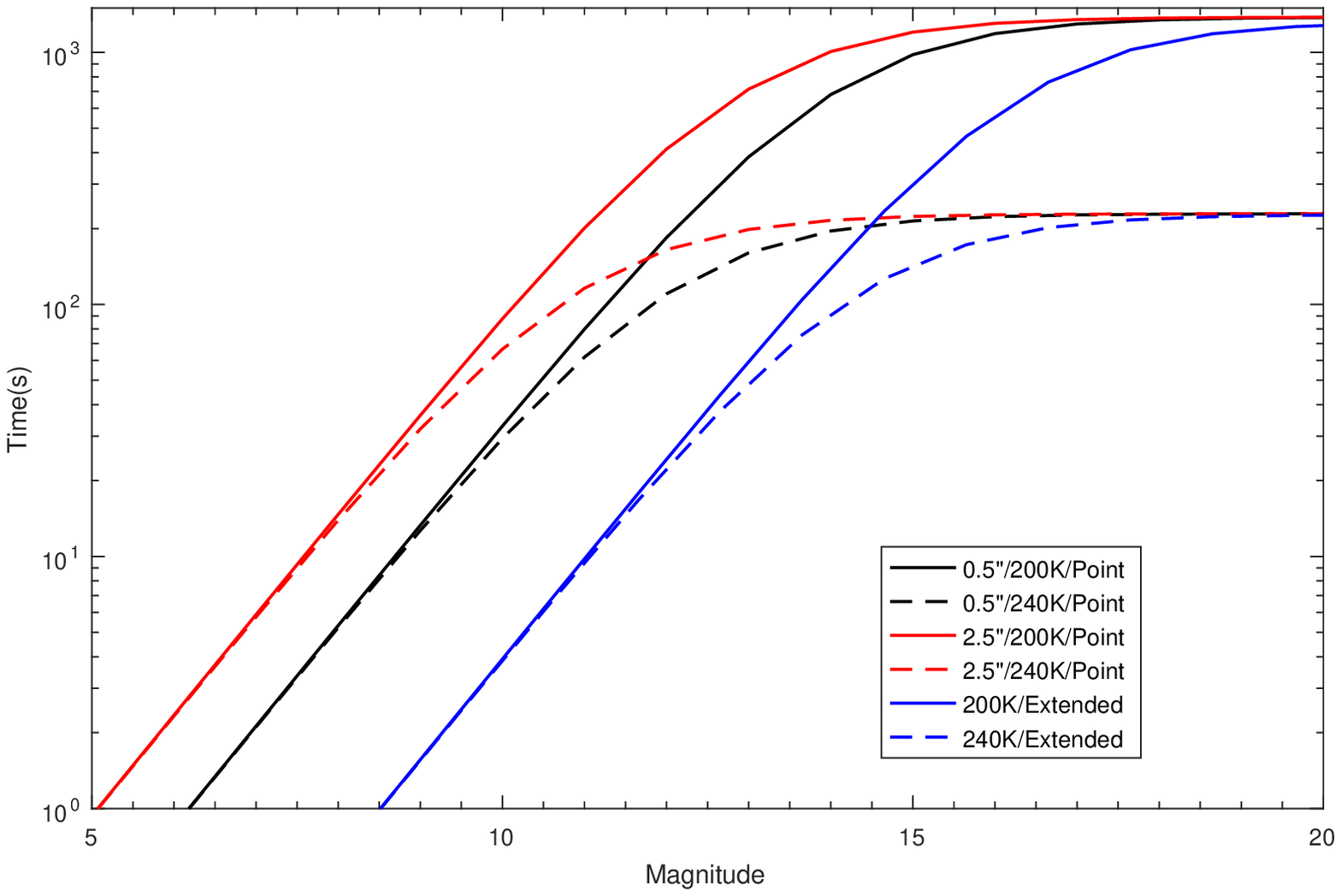}
		\caption{Maximum exposure time for a frame before saturation of the array occurs vs\@. Vega magnitude of point source.  Four different combinations of seeing and ambient temperature conditions are shown [$0.5''$ (black) or $2.5''$ (red) seeing;  200\,K (solid) or 240\,K (dashed) temperature], as indicated in the legend.  In addition, the saturation time for an extended source is shown in blue (with flux now in magnitudes per square arcsecond), for the same two background temperatures.  The three solid curves for the (cold) 200\,K  temperature asymptote to the same (long) fill time of 1,400\,s, while the three dashed curves for 240\,K saturate in about one-sixth this time.  Note that saturation is considered here as half-filling the wells in the detector. }
		\label{fig:MaxExposureTime}
	\end{center}
\end{figure*}

Figure \ref{fig:MaxExposureTime} shows the maximum exposure time possible before saturating the array {\it vs\@.}  object magnitude, for a range of different seeing and temperature conditions.  This time has been taken as that for half-filling the wells in the detector, so as to avoid any non-linearity issues that may arise from nearly-filled wells.  Saturation times are, of course, longer for bright stars in poor seeing as the stellar photons are spread over more pixels on the detector.    For faint stars, i.e.\ $\rm K_{dark} \gtrsim 14$\,mags., the background flux from telescope and sky determines the maximum integration time.  This can exceed 1,000\,s in the coldest conditions, about 6 times longer than in the poorest conditions.

The brightest star that can be measured depends on how quickly the array can be read out, in order to avoid saturation.  For a 2\,s read-out time, this equates to magnitudes of 7.0 and 5.9 for seeing conditions of $0.5''$ and $2.5''$, respectively, as can be read off the curves in Fig.~\ref{fig:MaxExposureTime}.  This upper limit is about 4 magnitudes brighter than for a 60\,s exposure time.

\begin{figure*}
	\begin{center}
		\includegraphics[width=2\columnwidth]{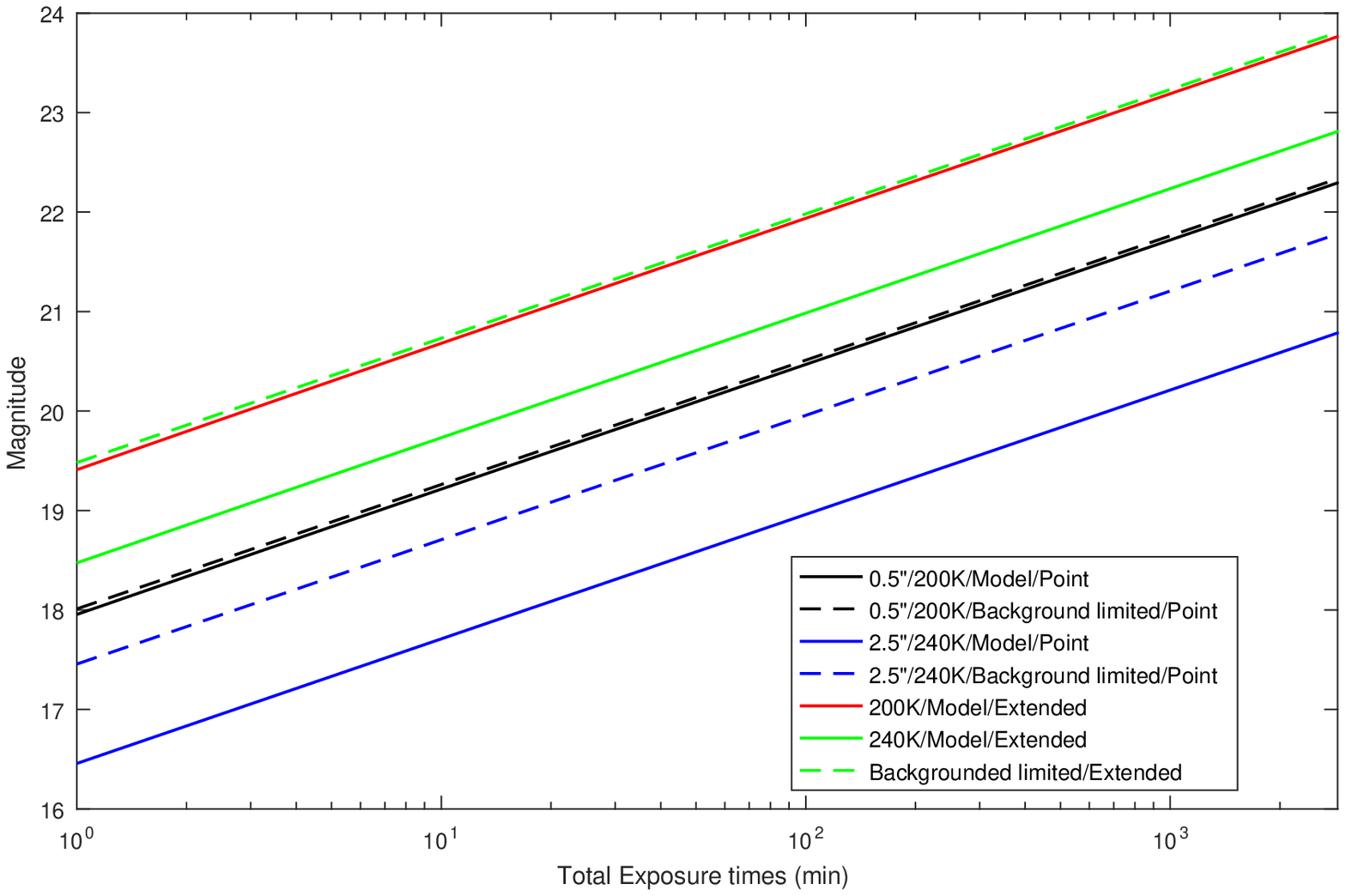}
		\caption{$3\sigma$ (i.e.\ S/N=3) sensitivity in Vega magnitudes for point sources vs.\ total exposure time in minutes, ranging from 1 minute to 2 days, in the two operating conditions considered ($\rm 0.5'', \,200\,K$ in black and $\rm 2.5'', \,240\,K$ in blue).  The corresponding dashed lines in the same colours are the limits if the observations were sky background limited for these seeing values.  For extended sources the flux unit is in magnitudes per square arcsecond, and is shown for ambient temperatures of 200\,K (red) and 240\,K (green).  Background limited operation for extended sources is shown by the green-dashed line. The total exposure time is considered to be made up of individual frames of 60\,s duration. }
		\label{fig:MagVsExp}
	\end{center}
\end{figure*}

Figure \ref{fig:MagVsExp}  shows the sensitivity vs exposure time for long integrations for the two extrema of weather conditions encountered at Dome A\@.  Here  we consider long exposures to be comprised of many frames of 60\,s integration time; if $30 \times 2$\,s were used then sensitivities would decrease by 0.4--1.0 mags, as shown in Fig.~\ref{fig:SNvsMag}. Typical point source $1\sigma$ sensitivities in 1 min, 1 hour and 1 day are  19.1, 21.4 \& 23.1 magnitudes in good conditions and 17.7, 19.9 \& 21.6 in poor conditions, as listed in Table~\ref{tab:sensitivity}.  The corresponding limits for extended sources are 20.6, 22.8 \& 24.5 magnitudes per square arcsecond in the coldest conditions (i.e.\ 200\,K) and  19.7, 21.9 \& 23.7 magnitudes per square arcsecond in the warmest (i.e.\ 240\,K).  The optimal performance is achieved when the telescope is limited by the sky background, with the sensitivity given by eqn.~\ref{eqn:Backgroundlimited}, as shown by the dashed lines in Fig.~\ref{fig:MagVsExp}.  As is apparent from the Figure, in the best conditions KISS should perform within 0.1 mags.\ of this limit.

Table \ref{tab:sensitivity} provides a summary of these model calculations for the camera performance under the two weather extrema.

\begin{table*}
\caption{Sensitivity for point source observations with KISS\@.}
\begin{center}
\begin{tabular}{@{}l c c c @{}}
\hline\hline
Parameter & Seeing $0.5"$ & Seeing $2.5"$ & Units  \\
                &  T=200K         &  T=240K         & \\
\hline%
Sky background & 16.9 & 16.9 & mags/arcsec$^2$  \\
                         & 100  & 100  & $\mu$Jy/arcsec$^2$  \\
Half-well fill time   & 1,380 & 229 & s  \\
1$\sigma$ sensitivity $30 \times 2$s  & 18.2 & 17.3 & mags.\\
1$\sigma$ sensitivity $1 \times 60$s  & 19.1 & 17.7 & mags.\\
1$\sigma$ sensitivity in half-well fill time  & 20.8 & 18.4 & mags. \\
1$\sigma$ sensitivity 1 hour  & 21.4 & 19.9 &  mags. \\
1$\sigma$ sensitivity 1 day & 23.1 & 21.6 & mags. \\
Saturation limit in 2s   & 7.0   & 5.9 & mags. \\
Saturation limit in 60s & 10.7 & 9.9 & mags. \\
\hline\hline
\end{tabular}
\end{center}
\label{tab:sensitivity}
Performance figures are calculated for a Teledyne chip with specifications as listed in Table~\ref{tab:specs}. Magnitudes are given in the Vega-system.
\end{table*}

\section{OPERATIONS}
\label{sec:operations}

\subsection{Remote Operation, Power and Communications in Antarctica}
\label{sec:comm}

In common with the optical AST3 telescopes, KISS will be at Kunlun Station at Dome A, which is only manned for a few weeks over summer. For the remainder of the year, KISS will be operated entirely remotely, with no possibility of on-site human intervention to correct any problems that may arise. Success therefore requires careful attention to reliability issues and redundancy of key components in a similar manner to a space mission.

The provision of power, a warm environment and satellite communications is provided by the PLATO (PLATeau Observatory) infrastructure that has operated reliably for over a decade in Antarctica \cite{2009PASA...26..415L,2009PASP..121..174Y}. PLATO currently provides 1\,kW of continuous power for a year from a combination of photovoltaic panels and diesel engines, but will need augmenting to supply the additional power requirements of the KISS instrument and telescope. Wind power is also a possibility, although Dome A is about the least windy place on earth \cite{2014PASP..126..868H}.

Satellite communications for Dome A is through the Iridium network, using two OpenPort modules---for redundancy, only one is used at a time---that provide 128\,kbps of bandwidth. Iridium modems, with 2,400\,bps bandwidth, are used as a backup means of communication for command and control. The amount of data that can be transferred from Dome A is limited by our budget, and is of the order of 100--200\,MB per month. This amount is small enough that we must process the data on-site, and only transmit back highly reduced data, thumbnails of images, and small portions of images at full resolution. An interesting option for the future would be a store-and-forward system based on a CubeSat satellite, which could increase the data transfer by a factor of 1,000, although it would require a tracking antenna that considerably increases the difficulty. For now, the raw data is stored on site and retrieved from Antarctica on disk drives during the annual traverses. We use hermetically-sealed helium-filled disk drives for reliability at the high altitudes at Dome A\@. Solid state disk drives are also an option, although we have experienced complete data loss if a drive is written to when it is too cold.

The remote operation of KISS has implications for the cryostat design, in that unless we provide a remotely operated vacuum pump, the cryostat needs to maintain a good vacuum in the event of interruption of power to the detector cooler. If the vacuum is not good enough, it may not be possible to cool the detector down again if it is allowed to warm up. We have four years of successful experience with operating cryostats and closed-cycle coolers at Ridge A, 150\,km from Dome A, as part of the HEAT/PLATO--R terahertz telescope \cite{2015ApJ...811...13B}; HEAT has recovered from a week-long power interruption, and was able to cool down again to 50\,K\@.

Perhaps surprisingly, over-heating can be a problem for instruments at Dome A due to the low air pressure, equivalent to an altitude of about 4,500\,m. This reduces the efficiency of air-cooling of electrical components, and requires the de-rating of power supplies by a factor of two. If electronic modules are allowed to cold-soak down to the ambient wintertime low temperatures of $-80^\circ$C, failures can occur, particularly of large encapsulated components where differential expansion is a problem.

\subsection{Environmental Protection}
\label{sec:protection}

A crucial aspect for any Antarctic telescope is to avoid ice formation or snow accumulation on any optical surface. The tube, including the primary and secondary mirror, will be sealed from the external environment using a window placed above the secondary mirror and a seal placed at the instrument flange. This enclosure provides for protection from snow and ice accumulation on the primary and secondary mirrors. An aggressive desiccant will be installed within the telescope enclosure to further prohibit ice formation on optical surfaces and an internal heater will be installed as a back-up to sublimate ice in case it forms. To avoid ice formation on the external surface of the telescope tube window, this optic will be coated with an indium-tin-oxide (or similar) layer that allows it to be heated. The telescope drive motors, gears, and bearings will be protected from the external air via a fabric tent covering the complete telescope. These components are all designed to operate, and have been tested, at the low temperatures experienced on the high plateau location.

\subsection{Telescope Control}
\label{sec:control}

KISS will be operated remotely in a similar manner to the first two AST3 telescopes. The regular survey is scheduled and managed automatically by an on-site observation control script. This script reads in a target list and calculates the observabilities of all sources in it. The priority of each target is preset to be``Normal'' or ``Special''. For targets labelled as Normal, parameters such as the required cadence, the airmass, distance to the Moon and distance to the limit positions of the mount are considered. A scanning sequence is then optimised by minimising the overall slew time. A target labelled as Special will be observed at the specified times if it is available. 

The target list is reviewed by the science team and sent to the on-site control computer at the beginning of the season. In general, it is not then changed to ensure completeness and consistency for the survey. However, Special targets can be added to the list as necessary.  Entirely new target lists can also be supplied if transients are identified for follow up.  The regular survey is then interrupted and the new observation set up and operated manually.

An observation log is also generated. Detailed information on the survey status, the instrument parameters and the environmental conditions are recorded in it. These are later used to evaluate the quality of the data during reduction. 

Sometimes operational failures may be encountered, such as exposure failures or pointing errors. The observation control script will then try to kill the previous action and restart the survey automatically. Meanwhile, an alarm is also sent to the telescope operators through emails and mobile phone messages, calling for human inspection of the observing program to ensure it it running correctly.

\subsection{Automated Pipeline Reduction}
\label{sec:reduce}

The data reduction pipeline is based on that currently operating for the first two (optical) AST3 telescopes at Dome A\@.  It comprises routines for image construction, astrometry and photometry. All images will be reduced automatically directly following their observations on-site. As discussed in \S\ref{sec:comm}, the limited communication bandwidth precludes sending raw images back from Antarctica for off-site processing.  Raw data and processed images need to be retrieved during the traverse the following summer for any specialised treatment required beyond the pipeline described below. 

Considering the limited on-site computation ability available, a basic image reduction will be performed for a standard time domain survey. This includes a dark-frame and bias/overscan subtraction and a flat-field correction. A master flat-field image will be constructed by a combination of twilight images obtained early in the season. A series of laboratory dark frames taken at various temperatures and exposure times will be used to model the corresponding dark frame for each science image. Median filtering of sequential frames will be used to determine sky frames for background subtraction.  Cleaned images will be passed to the photometry and astrometry pipeline.

For a standard domain survey image, the ``Sextractor'' \cite{1996A&AS..117..393B} package will be implemented to generate a raw source catalogue. Then the WCS information and zero-point magnitude will be determined by the ``SCAMP'' \cite{2006ASPC..351..112B} package, which matches the raw catalogue to the 2MASS database. With this information, an astrometry and photometry-calibrated catalogue can be generated by a second run of Sextractor. Finally, the whole catalogue, or a subset of pre-selected targets, will be sent back from Antarctica during good network conditions. In the case of poor network performance, light-curves of selected targets of interest will be generated on site and sent back for detailed de-trending and polishing. Additionally, the offsets on RA and Dec generated by SCAMP will be used to calibrate the tracking of the system. This is crucial for achieving high tracking accuracy and high photometric precision.

For the detection of transients additional sky templates need to be constructed for each target field during the best photometric conditions. These templates are cleaned and calibrated on both photometry and astrometry systems. The resultant standard templates are stored on-site. When better/deeper templates are taken,  the templates are updated. Before a transient survey image is passed to the photometry pipeline, its astrometry coordinates are calibrated by SCAMP and the corresponding template subtracted by a software package called ``Hotpants'' \cite{2000A&AS..144..363A}. Any emerging/brightening source or moving object will trigger an alarm and be passed to the photometry pipeline. Detailed photometry and astrometry information will be sent back and compared with the \emph{VizieR} database. Short light-curves (usually beginning with the date of the latest template) of new and interesting sources will be sent back for further study.  Differential photometry will be derived by comparison with bright stars in each frame. Flux calibration of the frames will rely on comparison with sky frames overlapping the ESO VISTA survey.

\section{SURVEY PROGRAMS}
\label{sec:survey}

\begin{table*}
\caption{Parameters for Sample Survey Programs}
\begin{center}
\begin{tabular}{@{}l c c c c  c@{}}
\hline\hline
Cadence & Samples    & Integration & Camera  & Area Covered & $3\sigma$ Sensitivity \\
              & per Winter & Time (s)      & Fields & Sq. Degrees   & Magnitudes \\
\hline
1 hour    & 2,000     & 1 minute      & 6 & 3.5                  & 16.5 - 17.9 \\
1 day      & 100        & 1 minute     & 144 & 85                & 16.5 - 17.9 \\
1 week   & 12           & 1 minute     & 1,008 & 600           & 16.5 - 17.9 \\
1 month & 4             & 1 minute     & 4,320 & 2,600       & 16.5 - 17.9 \\
1 day     & 100         & 9 minutes    & 16   & 9                 & 17.5 - 19.0 \\
1 week   & 12           & 9 minutes    & 112 & 66               & 17.5 - 19.0 \\
1 month & 4             & 9 minutes   & 480  & 280             & 17.5 - 19.0 \\
1 day     & 100          & 1 hour        & 2.4   & 1.4              & 18.7 - 20.2 \\
1 week   & 12           & 1 hour        & 17    & 10                & 18.7 - 20.2 \\
1 month & 4             & 1 hour        & 72    & 43                 & 18.7 - 20.2 \\    
\hline\hline
\end{tabular}
\end{center}
Cadence is the time between return visits to each field.  Camera Fields is the number of fields that would be surveyed in one winter season, with the corresponding areal coverage listed given the camera FOV. The range in $3\sigma$ sensitivity reflects the extrema in observing conditions at Dome A, as discussed in \S\ref{sec:sensitivity}.  Each survey is assumed to use 10\% of the available observing time over the Antarctic winter.  An integration time of $1 \times 60$\,s per frame is taken. Sensitivities if $30 \times 2$\,s were taken instead would be reduced by 0.4--1.0 mags.
\label{tab:surveys}
\end{table*}

In this section we take the sensitivities presented in \S\ref{sec:sensitivity} and discuss possible surveys that could be undertaken, in particular for exploring time variable phenomena. There are approximately 4 months of darkness each winter at Dome A, and this suggests a range of cadences for such programs, ranging from hours to monthly.  Naturally, the more frequently a field is re-observed, the greater the number of samples that will be obtained, but the smaller the total area of sky that can be surveyed.  In addition, if high sensitivity is required then the survey area will be further limited.  

We consider here surveys for cadences ranging from 1 hour to 1 month, and with integration times of 1 minute, 9 minutes (i.e.\ $3\times3$ frames) and 1 hour, as summarised in Table~\ref{tab:surveys}.  We have also assumed that each survey will be conducted for approximately 10\% of the available observing time over the winter.  In combination with the camera field of view (taken as $46' \times 46'$) and the integration time per field, this then determines the total areal coverage of a survey.  The integration time also determines the sensitivity.  For example, if a daily cadence is required a survey could cover 85 sq.\ deg.\ to a point source sensitivity limit of 17.9 mags with a 1 minute exposure.  However,  only 1.4 sq.\ deg.\ could be covered if 1 hour exposures were used in order to reach down to 20.2 mags.  In both these cases $\sim 100$ samples would be expected over the course of a season.

Furthermore, if the intention is to increase dynamic range across the field, or to include the brightest possible sources in a survey, then an integration time per frame of $30 \times 2$\,s can be used, instead of $1 \times 60$\,s.  This raises the brightest source measurable from 10.7 to 7.0 magnitudes, but results in a decrease in the detection limit of 1 mag.


\section{SCIENCE}
\label{sec:science}

The science mission of KISS is the widest field survey of the time varying Universe in the infrared. The VISTA Variables in the Via Lactea \cite{2010NewA...15..433M} survey has mapped a $\rm 560 \, deg^2$ area containing $\sim 3 \times 10^8$ point sources with multi-epoch near-infrared photometry.  Their surveyed area includes the Milky Way bulge and an adjacent section of the mid plane. KISS will venture outside the Galactic plane and bulge with a wide range of scientific objectives described below.

\subsection{Star Formation}

The motivation for studying star formation in the IR from Antarctica with KISS stems from three principal facets that improve the effectiveness of an observational program:
\begin{itemize}
\item The unique environmental conditions on the summit of the Antarctic plateau greatly improves capability.  The extreme cold results in low sky backgrounds in the $\rm 2.4\mu m \, K_{dark}$ window, $\sim 100$ times better than from good temperate sites.  Coupled with the long winter night and stable atmosphere, this facilitates sensitive, high cadence observations with precision photometry.  In turn, this allows comprehensive searches for variability and transient phenomena to be undertaken over extensive areas of obscured regions.
\item A small telescope with a large-array has a wide field of view.  Young star clusters are typically spread over tens of arcminutes of the sky, so this allows for them to be efficiently studied.  In particular, there are many clusters in the Galactic Plane that can be selected to well fit into a single pointing of the telescope.
\item Extinction is one-tenth that of the optical at 2.4$\mu$m, hence providing the ability to peer inside obscured molecular clouds where star formation takes place.  At shorter wavelengths, the interiors of star forming clouds remain hidden from view.
\end{itemize}

$\rm K_{dark}$ is the longest wavelength for which high spatial resolution, high sensitivity observations can be undertaken from a ground-based site. KISS thus opens a new regime for the time exploration of stellar variability associated with star formation in the infrared.

We now discuss some illustrative science programs that could be tackled with such a capability:

\subsubsection{Dynamics in Star Forming Cores}
Dynamical interactions and the decay of non-hierarchical multiple systems in the cores of star forming clusters might be found.  Such events may often be accompanied by stellar ejections, together with the formation of a tight binary, as suggested by the $\sim 30$\% of O stars that are runaways and the higher binary fraction in massive than in low mass systems \cite{1986ApJS...61..419G,2007ARAA..45..481Z,2008AA...489..105S}.  The runaways are presumed to have been ejected from their birthplaces at high velocities.  In OMC--1 a three-body encounter $\sim 500$ years ago has been hypothesised to have resulted in the formation of the ``bullets of Orion'' as well as the high proper motions of the three stars concerned \cite{1993Natur.363...54A,2011ApJ...727..113B}.  Proper motions studies of the bullets for over a decade clearly reveal their explosive origin \cite{2015AA...579A.130B}.  

The hypothesis to be investigated is that ejection, plus the formation of a tight binary, may be common in multiple-systems.  In this scenario massive stars form dense sub-clusters inside more sparsely populated clusters of lower-mass objects.  They might form non-hierarchical systems, whereby dynamical interactions between the members leads to the ejection of one at high velocity, with the simultaneous formation of tight binaries between other members.  While it would be fortuitous if such an ejection was to be caught in action, more likely would be the possibility of catching transits from the close passage of such stars in front of each another.  These must often occur as the orbital periods are short and the stellar photospheres can be extended.  Such events will only be observable in the infrared, however, since dust extinction renders these stars invisible in the optical.  Wide-field time domain studies, therefore, can search for such transiting events and so yield statistics on the existence of embedded massive, multiple systems during star cluster formation.

\subsubsection{Accretion Events in Star Formation}
Variability in the infrared fluxes from forming star clusters has been found to be surprisingly common.  For instance, in the nearby Rho Ophiuchi cloud 40\% of the embedded population exhibits such behaviour on timescales ranging from days to weeks \cite{2008AA...485..155A}.  Such variability is hypothesised to be associated with episodic accretion from a circumstellar disk, for instance outbursts (flares) might result from sudden accretion events.    As an example, in Cygnus OB7 variability was modelled as the accretion rate falling by 30 times, while the central hole in the disk doubled in size, over a 1 year period \cite{2013ApJ...773..145W}.  

There are two recognised classes of eruptive variables undergoing accretion: long duration outbursts over many years termed FUors (after FU Orionis), and shorter duration events (weeks to months) known as EXors (after EX Lupi).  These have been defined via their optical variability and only a few examples are known \cite{2010vaoa.conf...19R,2014prpl.conf..387A}.  

Searches for such events, in particular shorter period variability, has been hampered by a lack of capability in the IR for time-dependent surveys.  Two epochs in the near--IR were obtained during the UKIDDS Galactic Plane Survey (the UGPS) \cite{2014MNRAS.439.1829C}, but only finding an additional 4 examples arising from embedded sources.  

The first results focusing on IR variability from the much more extensive Vista Variables in the Via La\`{c}tea (VVV) survey \cite{2013BAAA...56..153C,2014Msngr.155...29H}  have recently been reported \cite{arXiv:1602.06267}, however.  They open up a much larger discovery space.  Around 7 large amplitude variables ($\Delta K_{s} > 1$\,mag) per square degree were found over a $119^{\circ}$ survey region, about half of which can be associated with embedded young stellar objects. After allowance for incompleteness, the actual source density for such variable may in fact be a factor of three higher.  Over 100 sources with variability characteristics of FUors and EXors have been identified, an increase by $\sim 5$ times over the number previously known.  

This provides a rich source list for more detailed study using a higher and more consistent cadence.  Interestingly, the most eruptive light curves, on the longest timescales, tend to be associated with the spectrally reddest objects; i.e.\ associated with the earliest stages of pre-MS evolution.  However, for shorter time intervals (less than a month) the greatest variability is associated with the more evolved class II YSO stage, rather than the class I stage found to be associated preferentially with longer period variables.  Monitoring of variability provides a tool to probe the accretion process itself, and the cycles of outburst that accompany it, so as to test between models for the process \cite{2016MNRAS.458.3299H}.  These can depend sensitively on the stellar structure and geometry in the accreting region.  These observations might be accompanied by ALMA imaging in the sub-millimetre of the disks around forming stars, in order to build a picture of the accretion process(es) in action. 

\subsubsection{Intrinsic Variability of Young Stellar Objects}
While fluctuations in the light curve may result from obscuration caused by other objects, it may also arise from variability of the young star itself.  On timescales of a few years Wolk et al., for instance, conclude that most of the young stellar population will be found to be variable in the near-infrared \cite{2013ApJ...773..145W}.  The causes of such variation are many, including those above (accretion, obscuration, transits).  However, it might also result from intrinsic variability -- perhaps from star spots changing the level of the emitted stellar flux, essentially changes in the level of magnetic activity \cite{2008AA...479..827G}.  

The nature of the variability can be used to probe the kind of star spot.  Periodic variability is thought to originate from cool magnetic spots that rotate with the star, whereas aperiodic variability is likely to be associated with flaring, irregular accretion or variations in circumstellar extinction.  Young stars are intrinsically more variable than main sequence stars like the Sun; however infrared observations are needed to study their variability when they are still obscured by the dust within their natal clouds.

Variability  might also be used to determine cluster membership from the background stellar population, in particular when the infrared excess that would arise from a warm disk that surrounds the young star is not seen.  Variability can therefore be used to help build a complete population census of a star cluster, beyond that provided by just the measurement of infrared excesses from disks.

\subsubsection{KISS Survey Program}
A suitable survey program for KISS would be to concentrate on observing star forming clusters along the southern Galactic plane.  Based on the results of the VVV survey about half the IR variable sources will be found in such regions.  These sources will not be evident in optical surveys, so this provides a new parameter space to study.  Young clusters will generally fit within the $46'$ FOV, so a 9-point jitter pattern would be used.  Thus in 9 minutes a $3\sigma$ sensitivity of $17.5 - 19.0$ mags.\ would be achieved for the Dome A extrema.  With daily cadence and a 10\% observing fraction, 16 clusters could be studied.  With a weekly cadence, this would increase to $\sim 100$.  

A suitable program would encompass both these cadences, concentrating on daily monitoring of the most interesting embedded clusters found in the VVV, including those with sources found to be continuously variable.  Source confusion in the heart of the Galactic plane will limit the sensitivity, so this monitoring would best be done in regions like Carina and NGC~3603 which are less crowded, being at longitudes $\sim 90^{\circ}$ from the Galactic centre.  This would be accompanied by a weekly monitoring of a larger sample of young clusters spread along the southern Galactic plane.

Based on the results of the VVV survey about 4 large amplitude ($\rm \Delta K_{dark} > 1$~mag) variables would be found per field brighter than $\sim 16.5$~mag.  Incompleteness in the VVV might raise this number to $\sim 10$. Since KISS will extend 2 mags.\ fainter, in fact considerably more large amplitude variables can be expected to be found.  Furthermore, KISS will also extend to brighter stars, with the saturation limit being $\sim 1$~mag.\ brighter than the VVV\@.

\subsection{Brown Dwarfs \& Hot Jupiters}
\label{sec:bds}

The $\rm K_{dark}$ window is where the spectrum of objects around 1,000--1,500\,K peak -- too cool for stars, but too hot for most planets. This is the regime of Brown Dwarfs and Hot Jupiters.  The latter are Jupiter-sized or larger planets.  They are in close orbit around their host stars and so heated by them.  While from comparison of their masses Brown Dwarfs look like the big brothers of giant planets, their origins may be very different. 

Brown Dwarfs are failed stars and are believed to be formed through the collapse of gaseous clouds, in the same manner as low mass stars \cite{2012ARA&A..50...65L}. In contrast, the formation of giant planets more likely follows a core accretion model \cite{2005Icar..179..415H}, in which the planet grows from a solid planetary core by accreting gas and dust from a surrounding circumstellar disk. 

A simple observational result seems to confirm their differences. Giant planets are relatively more common than Brown Dwarfs. It is estimated that 1--2\% of all stars may contain Hot Jupiters with orbital periods of less than 50 days \cite{2011ApJ...738...81W}. This rate increases to 14\% for orbital periods of up to 10 years \cite{Mayor11}.  The occurrence rate of Brown Dwarfs is lower than for both giant planets and low mass stars.  A possible dip occurs in the stellar companion mass histogram at around 40 Jupiter masses, known as the ``Brown Dwarf Desert'' \cite{2011A&A...525A..95S}. 

However, Brown Dwarfs are not readily distinguishable from giant planets if we only consider their masses or radii. There are planets which are smaller or larger, and lighter or heavier, than objects believed to be Brown Dwarfs. The Brown Dwarf Desert may in fact be an observational bias since they are much fainter than low mass stars and easily confused with giant planets. In fact, a debate about what is a planet began soon after the first Brown Dwarf was discovered and is still ongoing \cite{2006AREPS..34..193B,2015ApJ...810L..25H}. To contribute to resolving it, surveys are needed that are able to determine the true occurrence rate of Brown Dwarfs, especially in binary systems. This will not only benefit planet formation theory but also address some key issues in star formation research, such as the initial mass function for stellar mass objects \cite{2005PASA...22..199B}.

Stellar clusters are good targets for Brown Dwarf searches, since their occurrence rate may be strongly associated with the star formation rate. The youngest brown dwarfs and giant planets have dusty atmospheres \cite{2016arXiv160401411D}, which makes them bright in the K$_{\rm dark}$ filter. K$\rm _{\rm dark} - K_{\rm short}$ (2.1$\mu$m) should be one of the most sensitive measurements of the dustiness of their atmospheres, with similar sensitivity to the methane filters in the H--band (1.6$\mu$m). Targetting multiple clusters is then one of the most efficient ways of detecting Brown Dwarfs and Hot Jupiters. When they transit in front of their host star a $\sim 1$\% drop in flux of the host, over a period of a few hours, will be seen. To detect these tiny dips buried in the light curve, photometric measurements at the milli-magnitude accuracy level are required. Such a precision has been demonstrated to be achievable from Antarctica \cite{2014PASP..126..227F}.

Observations of transiting Brown Dwarfs and Hot Jupiters in the K--band have additional scientific value. Orbital inclination and physical radius can be determined more precisely, since stellar limb-darkening is significantly less pronounced at longer wavelengths. Most interestingly, the K--band is also open to secondary eclipse observations. The spectrum of a Hot Jupiter at 2.4$\mu$m will be much brighter than inferred from its black body temperature. So, when transiting a solar-type star, a secondary eclipse in the K--band of depth $\sim 0.1$\% is anticipated \cite{2005ApJ...626..523C,2015MNRAS.454.3002Z}. In addition, the secondary eclipse is even more pronounced for a transiting Brown Dwarf. 

Such measurements would provide a key part of the spectrum, and can be used to model the effective temperatures, Bond albedos and atmospheric physics. This would help us to distinguish Brown Dwarfs from giant planets and interpret their inner structure and formation processes.  Related work has been reported, e.g.\ HD209458b \cite{2005MNRAS.363..211S} and OGLE-TR-113 \cite{2007MNRAS.375..307S}, indicating the feasibility of this kind of observation. The sensitive $\rm K_{dark}$ window at Dome A and the long polar night will make KISS one of the best ground-based facilities for performing secondary eclipse observations of Brown Dwarfs and Hot Jupiters.

\subsubsection{KISS Survey Program}
The FOV of KISS is $46' \times 46'$ with a relatively small pixel scale $1.35''$.  This is very suitable for searching for transiting Brown Dwarfs and Hot Jupiters in open clusters along the southern Galactic plane. Multiple stellar clusters would be pre-selected depending on their stellar densities and distances. A trade off between total number of target clusters and sampling cadence needs to be made. The optimal balance is determined by ensuring enough exposure time to achieve at least 1\% photometric precision for each target and an overall cadence of less than 20 minutes. Since the typical transit duration of a Hot Jupiter is around 2--3 hours, a longer cadence would drop the detectability significantly given a fixed total observation period. Secondary eclipse observations would only be performed on very bright, high-value targets during their predicted transit windows. 

A target would be monitored for 4--6 hours to cover the whole transit event with a high sample cadence of 2 minutes. In a one minute observation 0.2\% photometry (see Fig.~\ref{fig:SNvsMag}) can be achieved for sources brighter than $10^{th}$ magnitude at K--band (for the extrema in conditions expected at Dome A, as discussed in\S\ref{sec:sensitivity}).  Ten one-minute images would be coadded together to achieve higher precision  $\sim 0.1\%$. 


\subsection{Exoplanets around M Dwarfs}
\label{sec:hjs}

In addition to finding individual objects in order to study their characteristic features, understanding of the statistical properties of the exoplanet population is essential for building a uniform and self-consistent picture of planet formation and evolution \cite{2015ARA&A..53..409W}. To date, most confirmed exoplanets have been found around solar-type stars (G Dwarfs). This could be an observational bias, however, since G Dwarfs are bright, quiet and slowly rotating, hence they are suitable for radial velocity follow-up measurements. In fact, main-sequence stars with spectral types later than M0 ($M_{\ast} < M_{\odot}$) outnumber G Dwarfs by an order of magnitude in the solar neighbourhood \cite{2006AJ....132.2360H}. Results from the Sloan Digital Sky Survey (SDSS) indicate that the Galaxy's stellar mass function peaks at $0.25 M_{\odot}$ \cite{2010AJ....139.2679B}, which corresponds roughly to M4 in spectral type.  40\% of stars fall below $0.25M_{\odot}$. We thus need to know how planets populate these smallest of stars before we can obtain a complete understanding of the exoplanet population in the Galaxy.

Furthermore, statistical studies on Kepler candidates and their host stars, which suffer less observational bias, indicate that smaller stars tend to have a higher probability of hosting small planets. The occurrence of Earth-- to Neptune-sized planets ($1 - 4 \, R_{\oplus}$) is higher towards later spectral types at all orbital periods. Planets around M Dwarfs occur twice as frequently as around G stars, and three times as frequently as around F stars \cite{2015ApJ...798..112M}. The correlation between planet occurrence and host mass reveals details of planet formation. For example, according to the core accretion paradigm \cite{2004ApJ...612L..73L}, Jupiter mass planets should be rare around M Dwarfs, since their less massive protostellar disks \cite{2005ApJ...631.1134A} are insufficient to feed giant planets. This hypothesis was supported by a deficit of Jupiters detected in radial velocity surveys \cite{2007ApJ...670..833J}. However, Hot Jupiters were then found around M Dwarfs a few years later by the Kepler satellite \cite{2012AJ....143..111J}. To date, only $\sim 100$ exoplanets around M Dwarfs have been found, mostly by Kepler\footnote{See http://exoplanet.org.}. 

We need more samples to complete the story of the stellar-mass-dependent occurrence rate of planets. By spanning a wide range of stellar masses, we may be able to draw an overall picture of how a planet's birth, growth and survival depends on its various environmental factors \cite{2007ApJ...669..606R,2009Icar..202....1M,2013ApJ...775...91B}.

Compared to gas giant planets, smaller rocky planets with moderate surface temperatures are particularly attractive for study as these are potential habitable worlds. An Earth-like planet usually has a physical radius no more than $2 R_{\oplus}$, which generates only a $\sim 0.04$\% flux drop when it transits its solar-type host. A photometric precision of $\sim 0.01$\% is required to reveal such planets in a transit survey. Furthermore, the typical orbit period of a habitable planet travelling around its G Dwarf host is about 1 year, which leads to a geometric transit probability close to zero. Given a system that is exactly coplanar to the observer, one needs to maintain this ultra-high precision for several years to ensure the coverage of multiple transit events.  This is impossible for ground-based surveys that suffer strong red noise from the atmosphere and sidereal aliasing.  Such issues are alleviated, but not removed, for observations from Antarctica.

Instead, M Dwarfs are preferred targets when searching for transiting Earth and super-Earth exoplanets in habitable zones \cite{2008PASP..120..317N}. The very low luminosities of M Dwarfs makes their habitable zones reside at much smaller orbital distances. For example, for an M5 Dwarf the habitable zone lies at around 0.08~AU, corresponding to an orbital period of 15 days. The geometric transit probability for a super-Earth exoplanet ($R_p \sim 2 R_{\oplus}$) inside its habitable zone is also raised from $\sim 0.5$\% (for the Earth-Sun system) to $\sim 1.6$\%.  Their small radii also significantly amplifies the transit depth from $\sim 0.04$\% to $\sim 0.5$\% \cite{2009IAUS..253...37I}. M Dwarfs are also much brighter at K--band than they are in the optical.  For targets of later than M5 or later spectral type and beyond the $\rm L \sim 13$ magnitude limit of TESS \cite{2015JATIS...1a4003R}, KISS has the potential to provide a unique survey for close in transiting terrestrial planets. A photometric precision better than 0.5\% is necessary for detecting  a super-Earth ($R_p \leq 2R_{\oplus}$) transiting a late M Dwarf ($R_{\ast} \leq 0.2 R_{\odot}$). The feasibility of detecting habitable Earth- and super-Earth planets has recently been demonstrated by MEarth \cite[GJ1132b]{2015Natur.527..204B} and TRAPPIST \cite[TRAPPIST-1b,c,d]{2016Natur.533..221G}.

\subsubsection{KISS Survey Program}
M Dwarfs are supposed to be more abundant than massive stars.  However, the nearby late M Dwarfs that are known are especially rare.  These are the best targets for exo-Earth searches and their distribution on the sky is rather sparse. For KISS, whose FOV is $46'\times 46'$, we expect less than one target per FOV\@. Thus, a blind survey is not a suitable search mode. Instead, a list of M Dwarfs will be chosen from the CONCH--SHELL catalogue \cite{2014MNRAS.443.2561G}, selected with radii $R_{\ast} \leq 0.3R_{\odot}$ and metallicity [Fe/H] $\geq 0$. Additional criteria are needed to ensure high photometric precision, e.g.\ bright K--band magnitude $M_K \leq 11$ and declination $\leq -60 ^\circ$ to be observable at high elevation.

In a one minute exposure the photometric precision is better 0.5\% (see Fig.~\ref{fig:SNvsMag}) for stars with $M_{K} \leq 11$. For bright targets, multiple short exposures (e.g.\ $30 \times 2$\,s) would be used to minimise saturation, with minor effect on the precision obtained (Fig.~\ref{fig:SNvsMag}). The typical transit duration of an Earth--M-dwarf system is about 1 hour. This sets the maximum cadence to be 10 minutes. Taking account of the slew time between targets (10--30 seconds), 6--8 targets could be monitored simultaneously.

\subsection{Terminal Phases of Stellar Evolution}

The terminal phase for intermediate mass stars is the Mira variable.  Heavy mass loss surrounds a $\rm \sim 10^5 \, L_{\odot}$ star, producing optically thick dust which obscures the process from view. The precursor optically thin stage is well studied, e.g.\ in the Magellanic Clouds \cite{1982ApJS...48..161A,1983MNRAS.202...59B,1998AA...338..592W}. Encoded in the time- and wavelength-dependent properties of these pulsating asymptotic giant branch (AGB) stars are the underlying fundamental parameters of mass, composition and evolutionary state \cite{2011ASPC..445...83I,2015MNRAS.450.3181V}.

\begin{figure*}
\begin{center}
\includegraphics[width=2\columnwidth]{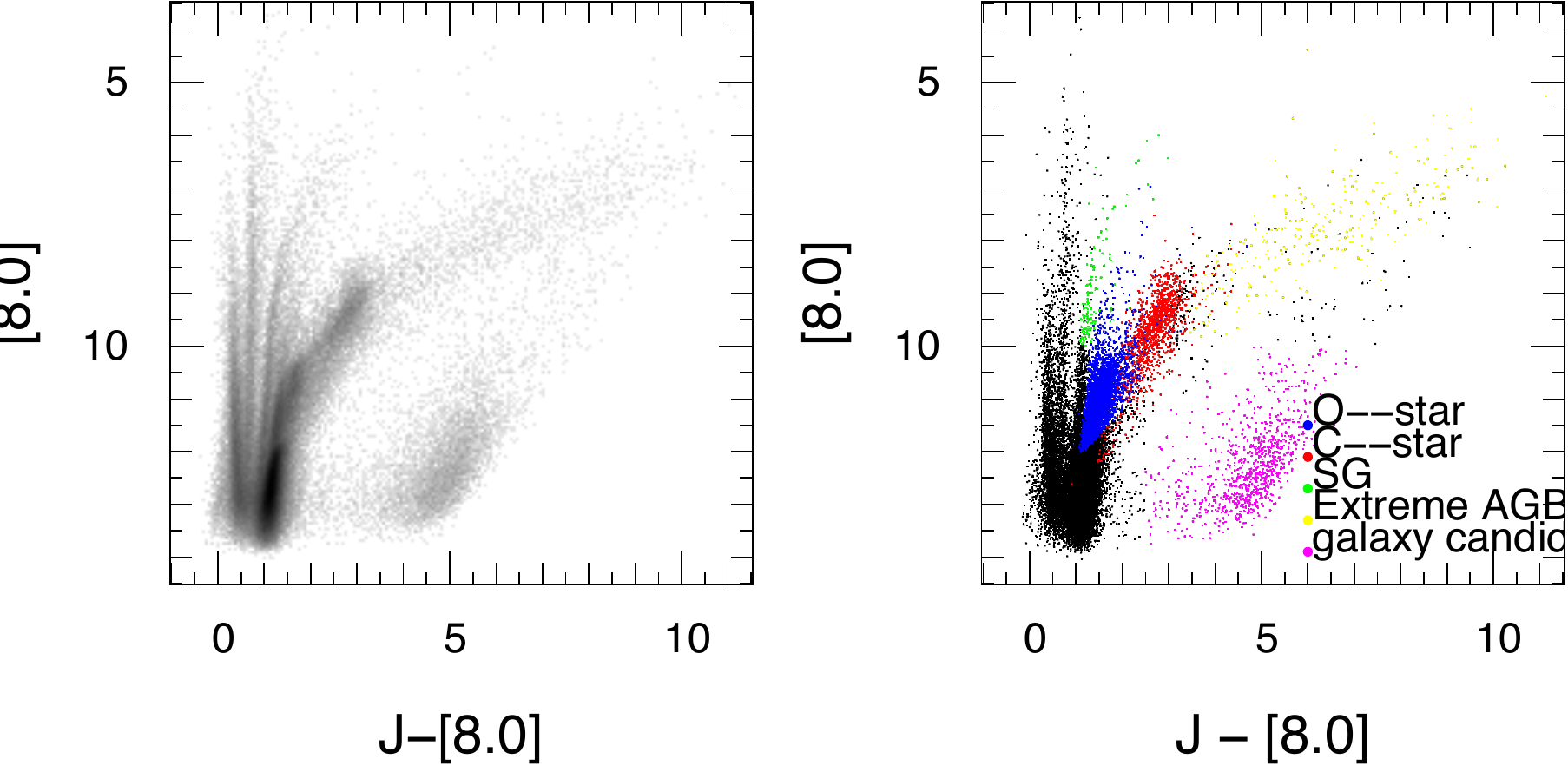}
\caption{Colour--magnitude diagram for extreme AGB stars from the SAGE survey (Woods et al.\ 2011), courtesy of Bob Blum.}
\label{fig:sage}
\end{center}
\end{figure*}

With $m_{\rm K} < 17.5$\,mags KISS can measure bolometric light curves for such stars in the LMC \& SMC, complementing the shorter wavelength work \cite{2013ApJ...779..167S}. Other complementary surveys to observe the AGB population of the Magellanic Clouds are the Deep Near Infrared Survey of the Southern Sky (DENIS) \cite{1994ApSS.217....3E}, the Surveying the Agents of a Galaxy's Evolution surveys with the Spitzer telescope for the LMC (SAGE--LMC) \cite{2006AJ....132.2268M} and the SMC (SAGE--SMC) \cite{2011AJ....142..102G}; see Fig.~\ref{fig:sage} \cite{2011MNRAS.411.1597W} for the corresponding colour-magnitude diagram for AGB stars. 

Mira period luminosity relationships have typically used V or I band periods combined with single epoch K band photometry. This is critical, because it is the mean luminosity of the variable that relates to the period, not the luminosity at single epochs. When infrared filter data are available, it has been difficult to collect sufficient data with a single instrument to uniquely identify pulsation modes \cite{2004A&A...413..903R}. KISS will enable a greater number of infrared photometric points for key nearby galaxies, including the Magellanic clouds and NGC 5128 (Centaurus A). When combined with optical photometry from surveys such as Skymapper and LSST, the degeneracy between initial mass and evolutionary state \cite{2015MNRAS.448.3829W} has the potential to be broken, enabling much more accurate distances and evolutionary states for individual objects.

\subsubsection{KISS Survey Program}
The area of the Magellanic Clouds is 100 square degrees. These fields should be observed at least twice a year for appropriate coverage of long period variables.  Mira variables are all brighter than the tip of the Red Giant Branch, which has an absolute K magnitude of about $-6.5$ (it is metallicity dependent). At the distance of the LMC/SMC, their magnitudes are brighter than 13, so that 1 minute exposures are easily sufficient.

For a study of the stellar populations of Centaurus A, several epochs are desirable targeting 0.1 mag precision at $\rm K_{\rm dark} =20$, extending the work of Rejkuba \cite{2004A&A...413..903R}.  This would require 6 hours of integration.  A minimum of two visits to the target per season would be needed.

\subsection{Fast Transients, Fast Radio Bursts, and Gravitational Wave
Sources}
\label{sec:dwf}

\begin{figure*}
\begin{center}
\includegraphics[width=2\columnwidth]{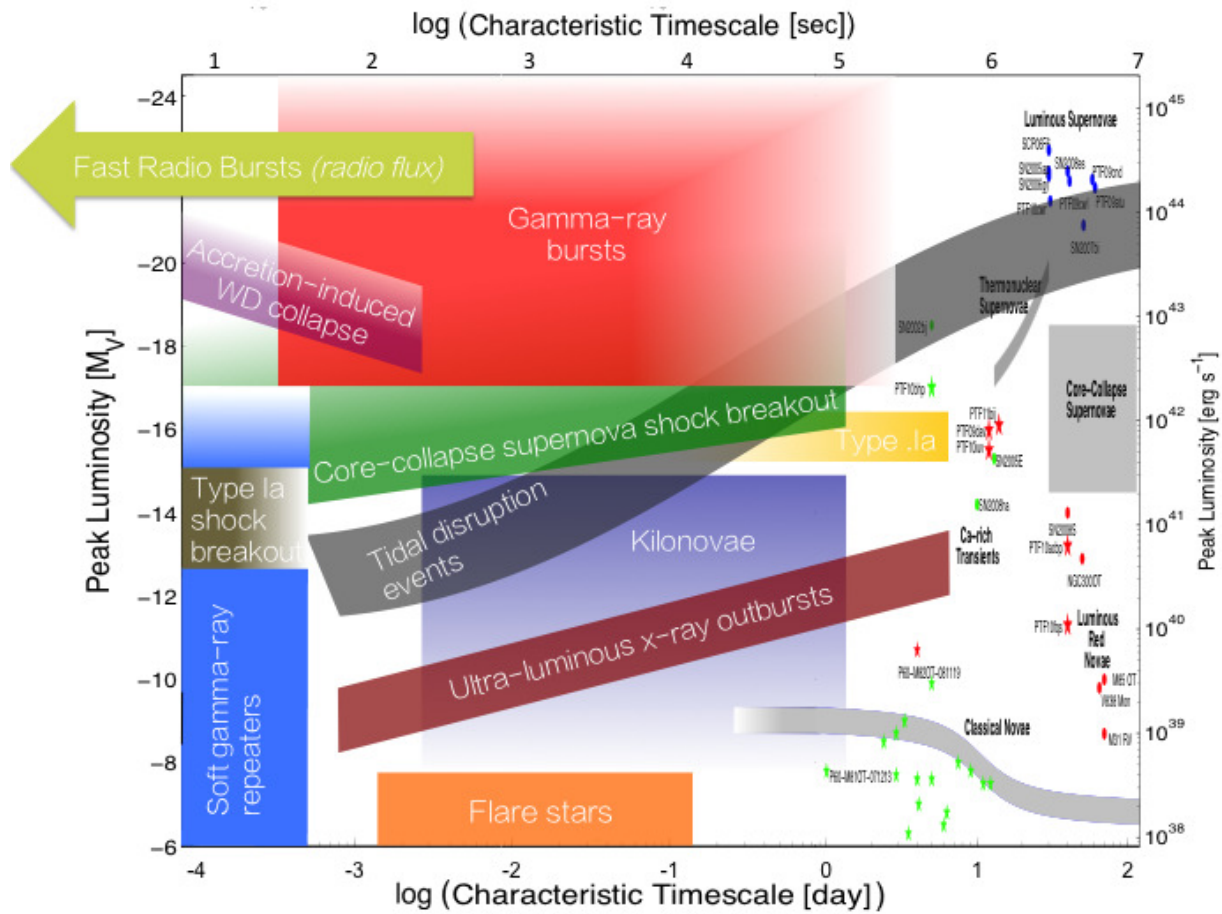}

\caption{Peak luminosity vs.\, characteristic timescale of the fastest (seconds-to-hours) theorised (and some observed) transient events, as well as conventionally observed transients with day to month timescales.  The figure is modified from Kasliwal et al.\ (2011) 
DWF performs simultaneous observations with multiple observatories from the radio to gamma-ray and is sensitive to many of these events at multiple wavelengths.  DWF core optical observations performed with DECam are sensitive to fast transient counterparts to very faint magnitudes.  DWF processes DECam images in minutes and identifies fast transients in real-time for rapid (minutes later) spectroscopic follow-up observations using 8m-class telescopes. Fast-cadenced, 20\,s DECam exposures reach to m$_g$(AB) $\sim$ 24 and nightly field stacks of m$_g$ (AB) $\sim 26$.  The images are capable of detecting nearly all events shown in the figure to $\sim 200$\,Mpc and events with peak magnitudes of M $\sim −16$ to M $\sim−18$ to $z\sim 1$. With continuous 20\,s exposures, DWF acquires detailed light curves of the ride and fade of most rapid bursts.  KISS will provide a key observational complement to DWF, by acquiring coordinated simultaneous infrared observations with the core DWF radio, optical, UV, X-ray and gamma-ray observations.  In addition, KISS will enable continuous 24-hour photometry, crucial to identify transient the timescales of $\sim 1$\,day, which are not possible with telescopes at any other latitude.  Finally, KISS will provide longer-term follow-up to confirm the fast transients, search for variability, and classify longer-duration events.  KISS observations are crucial for some gravitational wave sources, such as kilonovae, that are predicted to peak in the infrared and for events subject to Milky Way and host galaxy extinction. }
\label{fasttransients}
\end{center}
\end{figure*}

Recent wide-area transient surveys have identified thousands of events (e.g., the various classifications of novae and supernovae) that completely fill the luminosity vs.\ characteristic timescale parameter space from $\sim 1$\,day to $\sim 1$\,yr \cite{2011BASI...39..375K}.  However, there exists a large number of theorised, and a handful of serendipitously observed, faster evolving events occurring on millisecond-to-hour timescales (Figure~\ref{fasttransients}). Examples include fast radio bursts (FRBs) that are millisecond bursts in the radio that are likely of cosmological origin \cite{2013Sci...341...53T,2016Natur.530..453K}, supernova shock break-outs \cite{2008Natur.454..246S,2016ApJ...820...23G}, electromagnetic counterparts to gravitational waves \cite{2016PhRvL.116f1102A} such as kilonovae \cite{2015MNRAS.446.1115M}, gamma-ray bursts (GRBs), including `dark' and `bursty' events \cite{2008ApJ...678.1127K}, flare stars \cite{2005ApJ...621..398O,2012Natur.485..478M}, type .Ia supernovae \cite{2010ApJ...723L..98K} (supernovae with roughly one-tenth the timescale and luminosity as type Ia supernovae), and tidal disruption events \cite{2012Natur.485..217G}.  The millisecond-to-hours time domain has previously been little explored largely as a result of technological and observational challenges necessary for fast, real-time detection and study.

The Deeper, Wider, Faster (DWF) program \cite{cooke16} has overcome these challenges by:

\begin{itemize}
\item coordinating {\it simultaneous} observations using multiple, major (and, thus, sensitive) observatories at wavelengths from the radio to gamma-ray,

\item performing real-time supercomputer data reduction, calibration and analysis in seconds to minutes,

\item executing real-time computational and state-of-the-art visualisation technologies for real-time candidate identification,

\item initiating rapid response (minutes) deep spectroscopic follow-up programs using 8\,m-class telescopes, and

\item coordinating a global network of telescopes for short-to-long duration follow-up imaging.

\end{itemize}

\subsubsection{The Deeper, Wider, Faster (DWF) program} 

DWF organises and coordinates observations by a worldwide network of small to large telescopes to detect and study transients on millisecond-to-hours timescales.  As DWF is a `living' program, some facilities are added or changed over time as the program evolves and grows.  At its core, DWF performs simultaneous, deep, wide-field, fast-cadenced observations using the CTIO Dark Energy Camera (DECam) on the 4m Blanco telescope (optical), the NASA Swift satellite BAT (gamma-ray), XRT (X-ray) and UVOT (ultraviolet/optical) instruments, the 64m Parkes Observatory (radio), the 1 km Molonglo Observatory Synthesis Telescope (radio) and the Jansky Very Large Array (VLA) 25m (radio) telescopes.  Recently added to the simultaneous observations is the 0.68m Antarctic AST3-2 telescope (optical).  DWF also performs rapid response (within minutes) target of opportunity spectroscopic observations of fast transient candidates with the 8m Gemini-South telescope (optical/infrared), 0.3m NASA Swift UVOT (ultraviolet/optical) and proposed 8m VLT telescope (optical/infrared). Furthermore, DWF performs conventional (same night) target of opportunity spectroscopic observations with the 11m South African Large Telescope (SALT) and the 4m Anglo-Australian Telescope.  Finally, DWF coordinates a combination of interleaved and long-duration (weeks) photometry using the 1.3m Skymapper, 1m Zadko, 1m Mount Laguna Observatory and 0.68m Antarctica AST3-2 telescopes (all optical).

One of the primary goals of DWF is to provide simultaneous multi-wavelength coverage of FRB detections to search for shorter wavelength counterparts, localise the sources, confirm/refute their cosmological nature, study their host galaxies and help understand the mechanism(s) behind their energy. As a result, simultaneous data are acquired at radio, optical, UV, X-ray and gamma-ray wavelengths.  In addition, DWF has a Memorandum of Understanding (MoU) with the LIGO-Virgo Consortium and is notified of gravitational wave (GW) events.  Upon notification, DWF converts into a tiling mode and moves the core telescopes to cover the positional error ellipse of the GW event (depending on the ellipse location). DWF obtains temporal coverage of the event location at multiple wavelengths and to deeper magnitudes compared to other follow-up programs.  However, there is an important gap in wavelength coverage in the DWF network of telescopes at infrared wavelengths which KISS can fulfil.

Observations for DWF are classically proposed and scheduled, with coordinated runs currently occurring $2-3$ times a year at $\sim 1$ week each.  To enable simultaneous visibility for radio observations in Australia and optical observations in Chile, and to maximise time on source ($\sim 1-3$\,hrs), the physical limits of the telescopes require DWF target fields to reside between DEC $\rm \sim -30^{\circ} \, to \, \sim -80^{\circ}$.  As a result, the DWF fields are necessarily observable with KISS\@.  DWF receives immediate (seconds) FRB identifications from Parkes and/or Molonglo, bursts from Swift, and GW alerts from LIGO\@.  Apart from these rare, but high priority targets, the analysis of DECam optical images (candidates in minutes) dominate the transient detections.  The rapid (20\,s) DECam images are most sensitive to transients with m $\sim 16-21$ mags.  Fainter, slower evolving events can be detected in the stacked nightly, or semi-nightly images.  Although DWF performs real-time, fast data analysis and candidate identification, the program also conducts a comprehensive search for fast transients in the data after the observing runs.  Finally, some fast transients (e.g., supernova shock breakout, very early supernova detections, type .Ia supernovae) require long term (days/weeks) follow-up to identify and classify the events and to study their evolution.

\subsubsection{The role of KISS}
The geographical location, sensitivity, wavelength and temporal coverage of KISS make it an ideal observatory to complete the needs of DWF\@.  KISS observations can be coordinated with the DWF runs and can provide continuous, 24-hour photometric coverage during the observations.  The simultaneous and continuous coverage is crucial to obtain the light curves of events on minutes-to-hours timescales.  The ability of KISS to acquire complete, unbroken light curves of these events is extremely important and makes KISS unique.  Light curves for events with hours to $\sim 1$ day durations cannot be obtained by any higher latitude facility.  Fragmented photometry of such events would rarely provide sufficient data for confirmation and would not enable study and classification.

The infrared regime is vital for many fast transients detected by DWF\@. KISS is ideal for the detection and study of fast transients in fields near the Galactic plane, such as FRB fields, and are, thus, subject to heavy Galactic extinction at shorter wavelengths.  In addition, KISS will be highly effective in identifying and rapidly confirming flare stars, as the variation can result in negative flux in the infrared to account for the flare energy at other wavelengths \cite{1990IAUS..137..371R}. DWF detects early outbursts of supernovae and KISS will be key to obtaining their light curves over days for classification.  Finally, KISS will be a powerful facility to detect and confirm kilonovae (one aftermath scenario of the merger of two neutron stars), that are strong GW candidate sources, as they have hours-to-days timescales and are modelled to peak in the infrared \cite{2015MNRAS.446.1115M}.

KISS is projected to transmit source catalogues relatively quickly, which can be utilised in real-time during the DWF observation runs. In addition, archived KISS data will be highly useful for all transient detections and for the full analysis of DWF data performed after the runs.  Finally, all candidate GW source transients detected by DWF and KISS will be `reversed' searched in the LIGO/VIRGO databases.  Information of the time and location of the events enable higher sensitivity searches in the GW data.

\subsection{Supernova Searches}
\label{sec:sns}

\subsubsection{Type Ia Supernovae (SNe Ia) in the IR}
SNe Ia, after empirical calibrations of their luminosities \cite{1993ApJ...413L.105P}, are now the best standard candles to measure distances across the universe. Using SNe Ia, it was discovered in the late twentieth century that the expansion of the universe is accelerating \cite{1997ApJ...483..565P}. It is not yet known what causes this acceleration and this unknown energy is termed dark energy. Current observations supports the vacuum energy interpretation as the force behind the observed acceleration, yet other explanations are possible. If the vacuum energy interpretation is correct, and is constant over time, it violates fundamental physics and a new physical explanation is needed. On the other hand, the time varying nature of dark energy is hard to detect due to several systematic uncertainties in cosmological measurements. 

One way to minimise the systematics, which originated from the intrinsic luminosity scatter in SNe Ia, is to measure these objects in the IR\@. It has been shown that SNe Ia in the IR have the least scatter in the Hubble diagram \cite{2004AJ....128.3034K,2012MNRAS.425.1007B}, hence they provide better standard candles. Therefore, one important science goal of KISS is to discover SNe Ia in the IR\@. KISS will be able to discover hundreds of SNe Ia in the local universe and these will act as the low redshift anchor of the Hubble diagram. The high redshift part of the IR Hubble diagram can be filled with the SNe Ia from other future IR facilities, such as WFIRST\@.  SN Ia peak at K = 17.5 mag at 200 Mpc. There should be $\sim 200$/yr  SNe Ia detectable from Dome A with K $<$17.5 mag based on SDSS statistics \cite{2010ApJ...715.1021D}. KISS will also be able to undertake IR follow-up of optical SNe Ia that will be discovered from SkyMapper \cite{2007PASA...24....1K}. 

\subsubsection{Supernovae Buried in Starburst Galaxies}
Although SNe rates in starburst galaxies are higher, with current image subtraction technique in the optical, their discoveries are often hindered by their proximities to bright nuclei of host galaxies. In the IR, it is possible to see though to the central dusty obscuration of hosts and discover SNe. However, even with precision photometry, AGN variability (reverberation), if present in starburst galaxies, may affect positive SNe discoveries.

Therefore, an essential component of a project targeting SNe in the highly extinguished nuclei of star forming galaxies harbouring AGNs is a parallel survey program in the optical to a comparable depth as in the near-infrared. Such a parallel survey can be carried out on a small telescope, such as AST3--1 and AST3--2 at Dome A\@. Supernovae that are buried deeply inside the nuclei of AGNs have light curves that are fairly well understood. Optical light curves define the light curves of the AGN variability and so help to separate the supernova signal in the K--band from the variability associated with AGN activity.



\subsubsection{KISS Survey Program}

For the KISS SNe survey the aim is to make early SNe discoveries, i.e.\ soon after the explosion and before the peak in the light curve.  This requires a high cadence for the search, $\sim 1$ sample per day. Supposing 10\% of the winter observing time were devoted to such a search, then, based on the sample surveys in Table~\ref{tab:surveys}, in 60 seconds of integration we can cover $\sim 85 \deg^2$ with $\sim 144$ camera fields, and providing  $\sim 100$ samples per winter season. The $3 \sigma$ sensitivity achieved ranges from 16.5 - 17.9 mags.\ for the extrema in observing conditions experienced at Dome A, sufficient to make detections of SNe based on the SDSS statistics discussed above.  Optical imaging follow-up will be carried out by AST3--2, SkyMapper and La Silla Quest, and spectroscopic follow-up with the WiFeS instrument on the ANU 2.3\,m.  

\subsection{Reverberation Mapping of Active Galactic Nuclei (AGN)}

The region very close to the central supermassive black hole echoes its activity. In IR the dust morphology of the AGN is probed.  
The goal of infrared reverberation mapping is to characterise dust in the central disk or torus as a function of black hole mass and galaxy dynamics.
Monthly measurements of NGC 4151 have been modelled \cite{2013AA...557L..13S} by a static distribution of central ($\sim 0.1$pc) hot dust.  Acquisition of similar time domain data on a significant sample of nearby Seyfert galaxies is proposed as part of KISS\@. A suitable very nearby target is NGC1566 \cite{jud2016}.
The associated stars have central velocity dispersion measurable with the ANU's WiFeS and ESO's SINFONI instruments \cite{2015ApSS.356..347M}.  Together with the radius, this yields the mass of the black hole. The technique is as follows:

\begin{enumerate}
\item Monitor time variability of AGN at K$_{dark}$ (and at other wavelengths from Siding Spring Observatory).
\item Undertake a two-component decomposition of the SED, into a power law + black body.
\item Determine the temporal evolution of temperature + solid angle for the reverberating hot dust.
\end{enumerate}

\subsubsection{KISS Survey Program}
We plan to monitor the ten closest Seyferts in the Southern Hemisphere on a monthly basis.

\subsection{Gamma Ray Bursts (GRBs)}
\label{sec:grbs}


GRBs are the most violent stellar explosions in the Universe. At ultra-high redshifts (e.g.\ $z\sim20$) they require the infrared to be found. Models for WFIRST \cite{2014ApJ...797....9W} of the time evolution of GRBs from 1.5 to 4.4\,$\mu$m indicate that those at $z=20$ will be accessible in ${\rm K_{ dark}}$ from Antarctica. Currently, fewer than 10 GRBs have been found with redshift $6\leq z \leq 9.4$. High-$z$ GRBs are potential probes for the early Universe, exploring the first generation of stars (Pop III stars) and early cosmic star formation. With infrared spectroscopic observations, high-$z$ GRBs will provide opportunities to probe the re-ionization as well as cosmic metallicity evolution.
 
Intrinsic GRB and afterglow spectra are usually power-laws. Thus, combining ${\rm K_{ dark}}$ observations with simultaneous optical (e.g.\ $i$-band) measurements would easily identify whether the GRB is highly absorbed in the optical or not. If the optical emission from the GRB is heavily absorbed, then it may have been born in a dusty star-forming region.  Alternatively, its redshift is high. Dust obscuration and high-redshift are two possible origins of the so-called dark GRBs \cite{2001ApJ...562..654D}. In fact, the dust extinction in $37\%$ of GRB optical afterglows is $A_{\rm V}\sim 0.3 - 2.0$ mags, while $13\%$ of GRB afterglows suffer more dust absorption \cite{2013MNRAS.432.1231C}. Early follow-up observations after the GRB trigger in the ${\rm K_{ dark}}$-band will not only explore the properties of the infrared emission of their afterglows and hence the underlying physics related to the GRB outflows and circum-burst environments, but also initiate further spectroscopic observations with much larger telescopes if the joint ${\rm K_{ dark}}$ and $i$ band spectrum is unusual. The latter may lead to the discovery of high-redshift GRBs.
 
Compared with other wavelengths, such as gamma-ray, X-ray, optical, and even radio, the infrared follow-up is the most lacked band in GRB observations. Over the past two decades, many new features have been discovered, such as optical flashes during the prompt GRB phase, X-ray flares and plateaus in the early afterglow phase. As the infrared emission suffers no or little dust absorption, it is expected that the infrared observations will bring new information. One successful but rare example is the discovery of an infrared flash that accompanied the gamma-ray prompt emission of GRB 041219a \cite{2005Natur.435..181B}, which advanced our understanding of the physics of prompt GRBs. Infrared emission is not only expected during the prompt GRBs, but also in the afterglows, including the flaring stage and the plateau phase.
 
\subsubsection{KISS Survey Program}
As GRBs and afterglows are fast-fading transients, it would be better to do early (e.g., within 1 hour of the GRB trigger) follow-up observations with KISS\@. With contemporaneous observations at optical bands, real-time spectral analysis should be carried out to judge whether the GRB possibly originated at high-redshift.  If so further Target of Opportunity (TOO) spectroscopic observations would then be carried out with 10\,m class telescopes.

\subsection{The Cosmic Infrared Background}
\label{sec:cib}

$\rm K_{dark}$ is a choice wavelength for observing the first stars in the Epoch of Re-ionization (EoR), as the rest frame is the ultraviolet at $z=6$. Simulations \cite{2016MNRAS.457.1813F} show complementarity with observation of neutral hydrogen. The expectation is that the large-scale structure (LSS) in KISS will be anti-correlated with the LSS in MWA, for example.  The large scale power in the infrared background is not caused by the zodiacal light \cite{arxiv:1604.7291}.  

On smaller scales we have made a toy model for a $1000'' \times 1000'' $ square at $1''$/pixel 
with no background noise; the calculation stops at $z=10$.

The assumptions of the toy model are:
\begin{enumerate}
\item Galaxies aggregate according to Abramson et al. \ \cite{arXiv:1604.00016}.
\item The cosmic SFH is that of Gladders et al.\ \cite{2013ApJ...770...64G}.
\item Galaxy SEDs and M/L from Bressan et al.\ \cite{2012MNRAS.427..127B}\footnote{http://stev.oapd.inaf.it/cgi-bin/cmd}.
\item Standard cosmology with no extinction.
\item PSF=1 pixel.
\item The mean value of $\lambda F_\lambda$ is $\rm 22\, nW\,m^{-2} \,sr^{-1}$.
\end{enumerate}

\begin{figure*}
\begin{center}
\includegraphics[width=20pc, angle=-90]{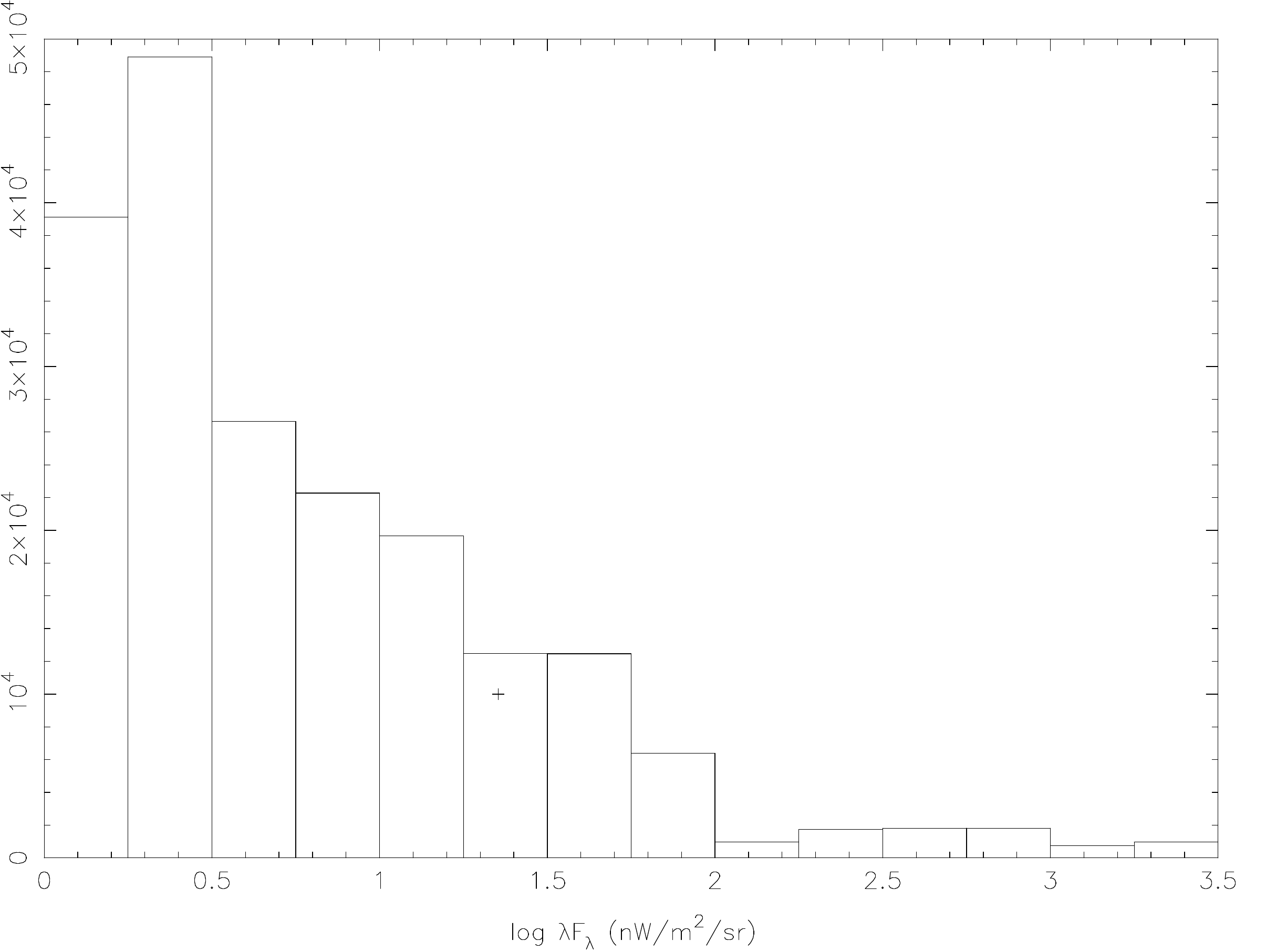}
\caption{Fluctuation spectrum of a toy model made with the assumptions itemized in \S\ref{sec:cib} and normalized to the DIRBE CIB, which is shown as a plus sign.  Number of pixels on the $y-$axis is plotted against the fluctuation in $\rm nW\,m^{-2} \,sr^{-1}$ on the $x$-axis.}
\label{fig:cib}
\end{center}
\end{figure*}

The fluctuation spectrum for the toy model is shown in Figure~\ref{fig:cib}. The normalization value of $\rm 22\,nW m^{-2} \,sr^{-1}$ was chosen from the DIRBE observations \cite{2004NewAR..48..465W}. The NICMOS deep field fluctuation spectrum for shorter wavelengths is discussed by Thompson et al.\ \cite{2007ApJ...657..669T}. The thermal background in low Earth orbit prohibited a K--band filter in NICMOS\@. More elaborate models for the KISS fluctuation spectrum will constrain star formation at high redshift. A high signal to noise deep field will be required to detect the fluctuation spectrum as the Kunlun Station winter background at $\rm K_{dark}$ is 100 $\mu$Jy per arcsec$^2$, which is 240 times the DIRBE CIB\@.

\subsubsection{KISS Survey Program}
We plan to evolve a strategy to obtain the deepest possible field to probe the cosmic infrared background, in combination with high cadence measurements obtained from other survey programs.

\section{CONCLUSION}
\label{sec:conclusion}
KISS will provide the first comprehensive exploration of the time varying Universe in the infrared at the longest wavelength where deep, high resolution observations can be conducted from the ground, $2.4\mu$m.  This will also be a demonstrator for the capabilities of large infrared telescopes on the Antarctic plateau that are able to take advantage of the exceptionally cold, dry and stable conditions for sensitive  observations of the cosmos in these wavebands.

%
%



\begin{acknowledgements}
We acknowledge the Australian Research Council for funding of the KISS camera under the Linkage Infrastructure Equipment Facility (LIEF) scheme, grant number LE150100024.  Useful conversations were held with Maren Hempel and Phil Lucas which helped guide the survey planning.  We are also grateful for the support of ACAMAR, the Australia-ChinA ConsortiuM for Astrophysical Research, which helped stimulate the work presented here by bringing the participants together.  

JC acknowledges the Australian Research Council Future Fellowship FT130101219.  Parts of this research were conducted by the Australian Research Council Centre of Excellence for All-sky Astrophysics (CAASTRO), through project number CE110001020.
\end{acknowledgements}




\end{document}